\documentclass[]{spie}  

\usepackage{amsmath,amsfonts,amssymb}
\usepackage{graphicx}
\usepackage[colorlinks=true, allcolors=blue]{hyperref}

\title{High Contrast Demonstrations of Novel Scalar Vortex Coronagraph Designs at the High Contrast Spectroscopy Testbed
}

\author[a]{Niyati Desai}
\author[a]{Jorge Llop-Sayson}
\author[a]{Nemanja Jovanovic} 
\author[b]{Garreth Ruane}
\author[b]{Eugene Serabyn}
\author[b]{Stefan Martin}
\author[a]{Dimitri Mawet}
\affil[a]{Department of Astronomy, California Institute of Technology, 1200 East California Blvd., Pasadena, CA, 91125}
\affil[b]{Jet Propulsion Laboratory, California Institute of Technology, Pasadena, CA 91109, USA}

\authorinfo{Send correspondence to ndesai2@caltech.edu}

\begin{document} 
\maketitle

\begin{abstract}

For direct imaging of exoplanets, Scalar Vortex Coronagraphs (SVCs) are an attractive alternative to the popularly used Vector Vortex Coronagraphs (VVCs). This is primarily because they are able to induce the same phase ramp regardless of the incoming light's polarization state. We tested a set of stepped SVC staircase masks in the Exoplanet Technology Laboratory (ET Lab) at Caltech on the High-Contrast Spectroscopy Testbed (HCST). Here we present some preliminary findings of their starlight suppression ability, achieving raw contrasts on the order of $10^{-5}$ for 7 to 9 $\lambda$/D. We also characterized their chromatic performance and performed wavefront control to achieve preliminary contrasts on the order of $10^{-7}$ with EFC. These initial experimental results with SVCs have shown scalar vortex technology has a great potential for future exoplanet direct imaging missions.

\end{abstract}

\keywords{High contrast imaging, instrumentation, exoplanets, scalar vortex coronagraph}

\section{INTRODUCTION}
\label{sec:intro}  

High contrast direct imaging of exoplanets has proven to be an important future direction for research because of its potential to image Earth-like planets close enough to their host stars to be in the habitable zone. Flagship mission concepts HabEx\cite{HabExReport} and LUVOIR\cite{LUVOIRReport}, under consideration for the upcoming Astro2020 Decadal Survey, are driving motivators for advancements in coronagraph instruments as either one would be the very first direct imaging mission dedicated to Earth-like exoplanet imaging and characterization. Both concepts currently have mission requirements for a coronagraph which achieves contrasts on the order of $10^{-10}$ over a 20$\%$ bandwidth.

Various coronagraph instruments aim to optimize performance by considering the trade-offs between inner working angle, contrast, throughput, and sensitivity to low-order aberrations. Vortex coronagraphs\cite{Mawet2005, Foo2005} are so far one of the most promising options because they have demonstrated a desirable balance between these parameters by allowing for high throughput of planet light along with a small inner working angle and low sensitivity to low-order aberrations. Both the HabEx and LUVOIR mission concepts currently have vortex coronagraphs chosen as their baseline coronagraph to accomplish the science objective of exoplanet direct imaging\cite{HabExReport, LUVOIRReport}.

\section{VORTEX CORONAGRAPHS}
\label{sec:design&fab}
There are two main flavors of vortex coronagraphs: scalar and vector. A vortex coronagraph diffracts the starlight that is centered on the mask away from the center to be blocked by a Lyot stop downstream in the pupil plane. Both effectively turn the incoming starlight into an optical vortex, but in fundamentally different ways. There has been significant progress made in the design, fabrication and testing of vector vortex masks in the last decade however not much similar exploration of scalar vortex masks had been carried out thus far (see section \ref{subsec:VVC}). This work builds directly upon early literature by Swartzlander 2006\cite{Swartzlander2006} and more recent literature by Ruane et al. 2019\cite{Ruane2019} which both propose the theoretical and simulated potential that scalar vortex coronagraphs (SVCs) offer for advancing high contrast imaging of exoplanets. Although they offer a solution to creating an achromatic SVC through the stacking of two or more scalar vortex masks, the focus of this paper will be on the proof-of-concept testing of a prototype monochromatic SVC.

\subsection{Vector Vortex Achievements and Limitations}
\label{subsec:VVC}

Vector vortex coronagraphs (VVCs)\cite{2010ApJ} have thus far already been very successful in exoplanet direct imaging from the ground. Advancements in VVC design and fabrication have led to their use at Keck\cite{Wang2020,LlopSayson2021} and the VLT\cite{Wagner2021}. VVCs have already reached raw contrast levels of $10^{-8}$ at 10\% bandwidth in testbed environment\cite{Serabyn2019}. Further progress is currently being explored in terms of this coronagraph's limitations\cite{Ruane2020}.

A vector vortex coronagraph is essentially a half-wave plate whose fast axis varies azimuthally. VVCs achieve starlight suppression by imprinting a continuous either positive or negative phase ramp on the circularly polarized components of the wavefront. Equation \ref{eq:jones} shows the Jones matrix in the circular polarization basis which demonstrates how the wavefront is modified by the vector vortex focal plane mask.

\begin{equation}
\label{eq:jones}
\mathbf{M}_{\circlearrowright}=c_{V}\left[\begin{array}{cc}
0 & e^{i l \theta} \\
e^{-i l \theta} & 0
\end{array}\right]+c_{L}\left[\begin{array}{cc}
1 & 0 \\
0 & 1
\end{array}\right]
\end{equation}

Here the first term indicates that a vortex phase ramp, exp($\pm il\theta$), is applied to each polarization component, where the sign of the vortex ramp depends on whether the incident circular polarization is right or left handed. The second term is the stellar leakage component due to imperfect retardance. Imperfect retardance simply means the beam is passing through the vector vortex without picking up a vortex phase. The simple phase flip shown in the off-diagonal terms of the Jones matrix necessitates either splitting the polarization components, or filtering one out, effectively isolating each of the two directions.

Because of these split direction phase ramps, current VVC implementations involve introducing several additional optics, which not only require further precise alignment, but also add to the cost and complexity of the optical setup. This dependence on polarization is the primary disadvantage of VVCs because focal plane wavefront control algorithms such as electric field conjugation (EFC) can only sense one phase at a time, and generally can’t come up with a single deformable mirror (DM) solution for these induced spirals of opposite signs.

One solution to this requires introducing a polarizer and an analyzer at the front and end of the light path, ensuring that only one circular polarization reaches the focal plane and enters the vector vortex mask (see Llop-Sayson et al. in these proceedings). Another solution is to duplicate the entire instrument for each of the positive and negative phase-ramped polarizations by way of a polarizing beamsplitter. Although polarization filtering helps cut out the leakage term and isolate only one of the spiral phase ramps, it also effectively cuts the total throughput per polarization channel in half. 

\subsection{Scalar Vortex Potential}

The primary motivation behind exploring scalar vortex coronagraphs is to circumvent the polarization dependent limitation imposed by vector vortex masks. A scalar vortex mask achieves this by fundamentally imprinting a spiral phase pattern differently than a vector vortex mask. Essentially, a scalar vortex mask is a phase-only focal plane mask which varies in either surface height or in index of refraction with respect to azimuth. Equation \ref{eq:transmissivity} shows the complex transmissivity of an SVC:
\begin{equation}
\label{eq:transmissivity}
    t=\exp \left(i \frac{4 \pi}{\lambda} h(\theta)\right)
\end{equation}
where $h(\theta)$ is the surface height of the mask as a function of angle and $\lambda$ is the wavelength of the incident light. For the simple scalar vortex mask design where the thickness increases with azimuthal angle, equation \ref{eq:height} describes the surface height\cite{Ruane2019}:
\begin{equation}
\label{eq:height}
    h(\theta)=\frac{l_{0} \lambda_{0} \theta}{2 \pi\left(n\left(\lambda_{0}\right)-1\right)}
\end{equation}
Here $l_{0}$ is the designed charge of the optical vortex, or the number of times the phase wraps around the center of the spiral, $\lambda_{0}$ is the centrally designed wavelength, and $n(\lambda_{0})$ is the index of refraction of the mask. To understand the chromatic behavior of the scalar vortex mask, notice the dependence on wavelength in the surface height equation. So at a wavelength of $\lambda$, the phase shift of the transmitted light is $l_{0} \theta$. For light at any wavelength other than the designed central wavelength $\lambda_{0}$, the charge of the vortex will vary. Even if $n(\lambda)$ is approximately constant across the bandwidth, the charge goes like $l (\lambda) = l_{0} \lambda_{0} / \lambda$.

This monochromatic property intrinsic to single-layer SVCs is the primary limitation keeping them from currently being a viable coronagraph choice which meets the performance (bandwidth) requirements for space-based direct imaging. Refer to Ruane et al. 2019 \cite{Ruane2019} for a more thorough analysis on the theory behind the SVC.

\section{SCALAR VORTEX MASK DESIGN \& FABRICATION}
\label{sec:design&fab}

Several scalar vortex mask designs proposed by Ruane et al. 2019\cite{Ruane2019} were considered for fabrication and testing.  Designs ranged from azimuthal cosine masks, to stepped masks with various thicknesses. See section \ref{sec:future} of this paper for other alternatively fabricated scalar vortex mask designs which are in the process of being tested (fabrication techniques include ultrafast laser inscription and nano-post technologies). Stepped masks can vary in thickness while still imparting the same phase delay because the ramp can be made up of a combination of ramps, where each ramp sector has discontinuous steps equivalent to an integer multiple of 2$\pi$ radians. The number of discontinuities in the mask thickness is referred to as pitch multiplicity\cite{Ruane2019}. One of the primary motivations to vary pitch multiplicity is to reduce the overall thickness of the spiral phase plate and the height of the discontinuities. This would not only ease manufacturing requirements, but also ease alignment of the focal point exactly midway between the front and back faces of the focal plane mask.

\begin{figure} [ht]
\begin{center}
\begin{tabular}{c c} 
\includegraphics[height=5cm]{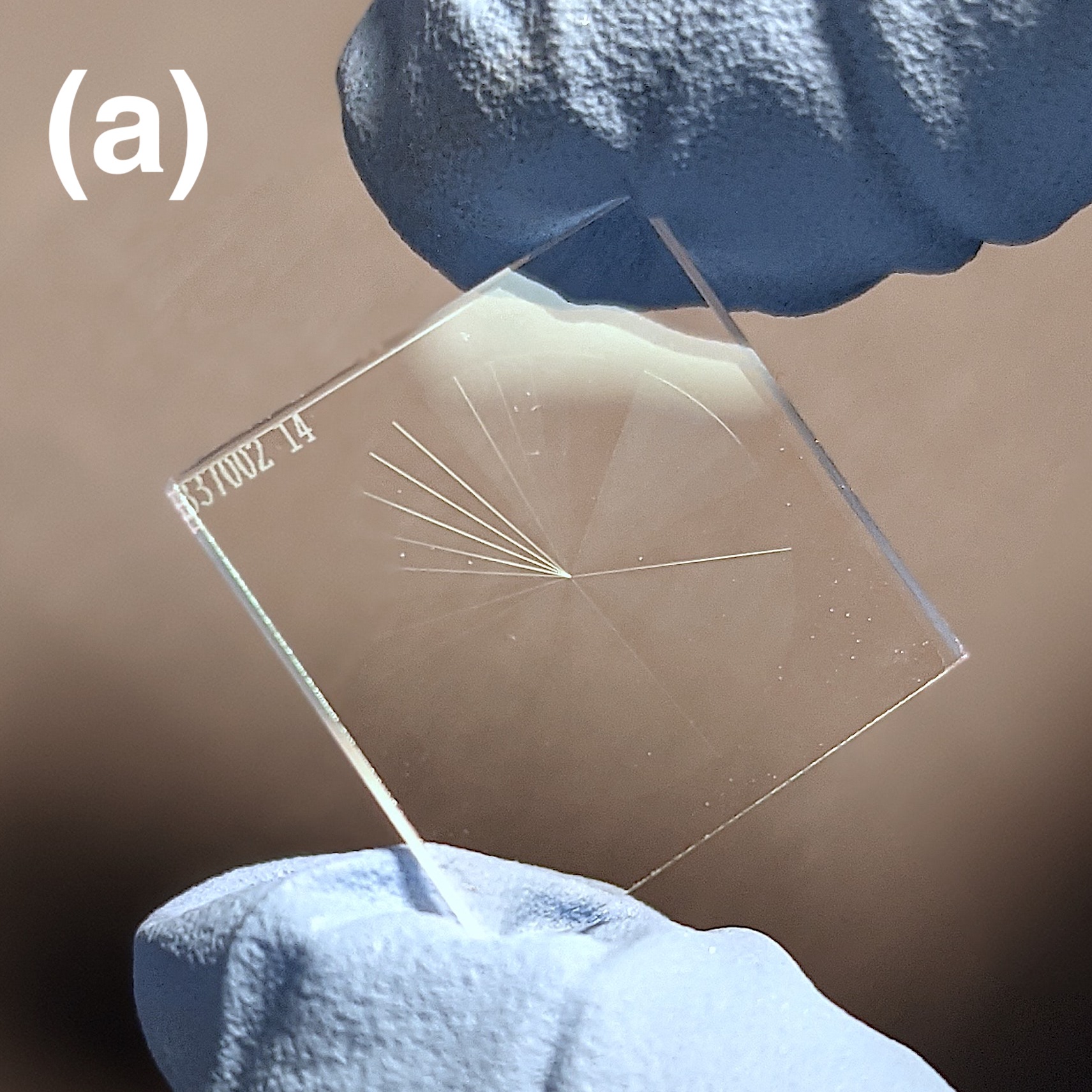}
\includegraphics[height=5cm]{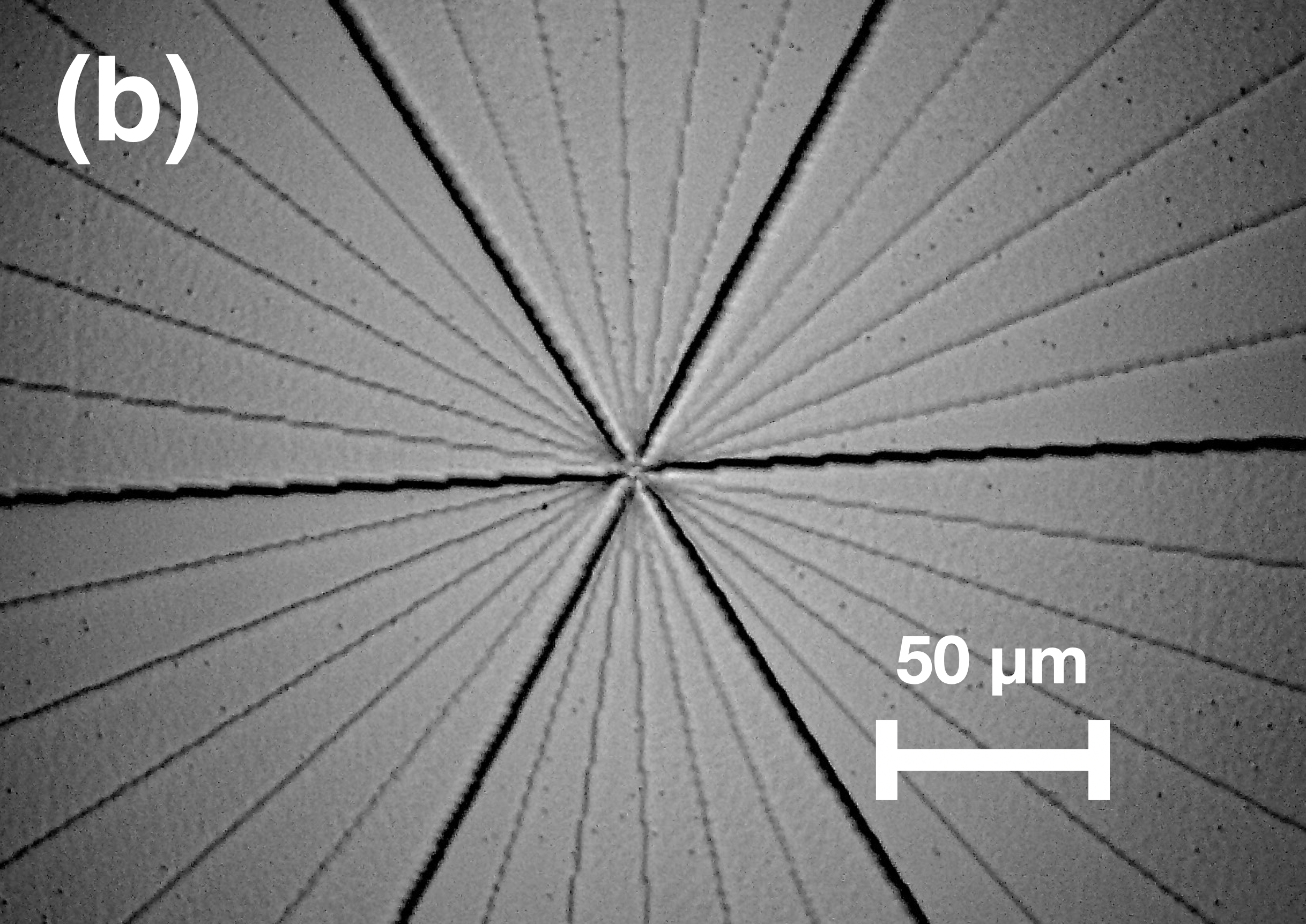}
\end{tabular}
\end{center}
\caption[fig:svcimages] 
{ \label{fig:svcimages} 
(a) Picture of prototype stepped scalar vortex mask. (b) Microscope image of central defect at 904x magnification.}
\end{figure}

The scalar vortex mask tested here was a stepped scalar vortex mask fabricated by Zeiss and designed to have a charge of 6. It had a pitch multiplicity of 6 and each of the 6 discontinuous sectors was a ramp also made up of 6 steps (see Fig.~\ref{fig:svcimages}). The mask was fabricated on fused silica and created for a central wavelength of 775nm. The square silica sample is 15mm by 15mm in size and the actual spiral phase plate region is circular with a radius of 13mm and centered in the square.

An important measure of the quality of a focal plane mask, particularly vortex masks, is the size of the central defect. Microscope measurements revealed the central defect on the mask to be no larger than approximately 8.2 microns in diameter. Furthermore, the ridging along each of the steps visible in the microscope image (Fig.~\ref{fig:svcimages}) was measured to be less than approximately 11 microns. For the high contrast testbed used in this study, it is imperative that the central defect be small relative to the beam/spot size. Since the focal ratios of the optical setup here use large focal ratios (F/\# of $\sim$30), this measured central defect is negligible.

\section{LABORATORY SETUP}
\label{sec:design&fab}

\begin{figure} [ht]
\begin{center}
\begin{tabular}{c} 
\includegraphics[height=5.5cm]{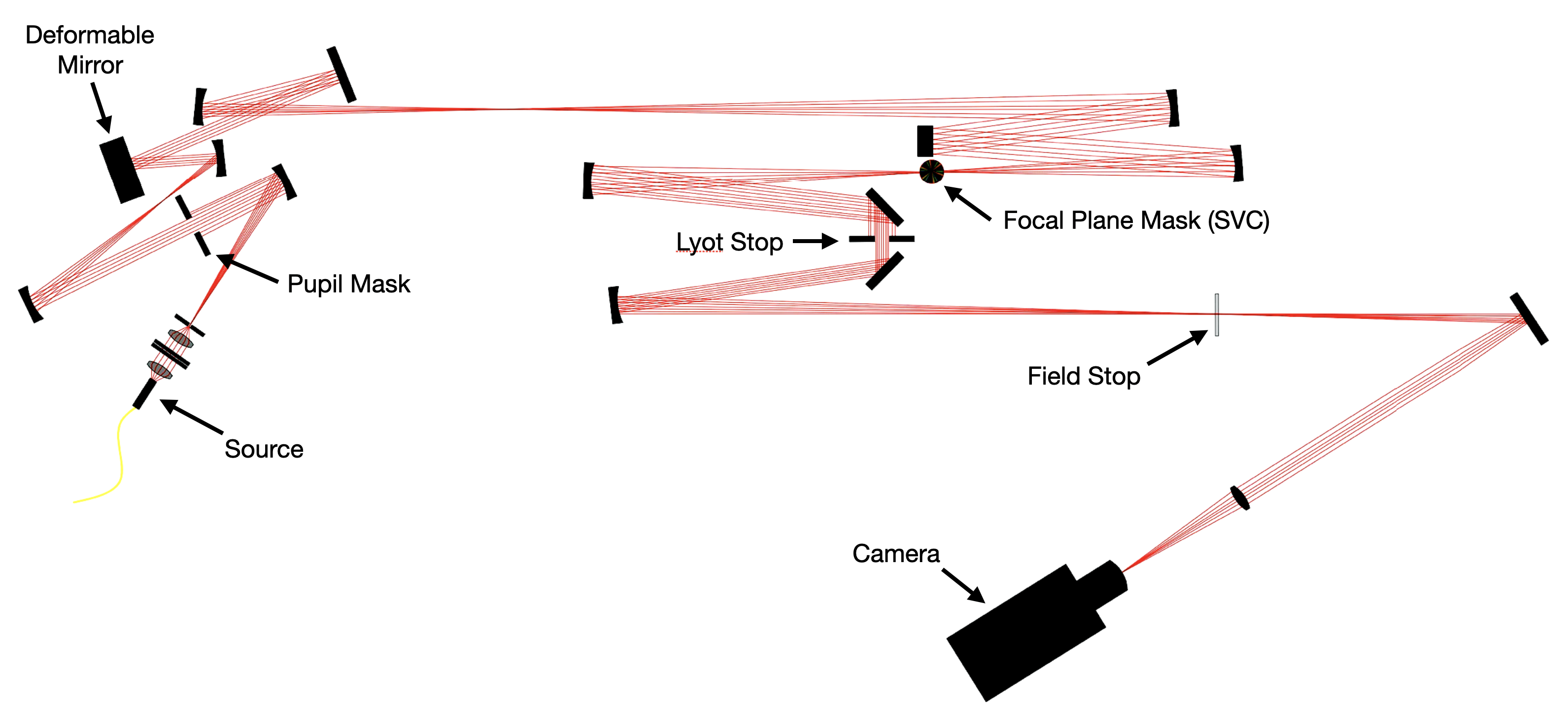}
\end{tabular}
\end{center}
\caption[fig:hcst] 
{ \label{fig:hcst} 
Layout of the High Contrast Spectroscopy Testbed for demonstrating the scalar vortex coronagraph. Refer to Llop-Sayson et al. 2020a \cite{Llop-Sayson_SPIE2020} and 2020b\cite{AVC2020} for details about HCST testbed design and setup.}
\end{figure}

Here we present the first laboratory demonstration of an SVC concept using the High Contrast Spectroscopy Testbed (HCST) in the Exoplanet Technology Laboratory (ET lab) at Caltech. The layout for the HCST is shown in Fig.~\ref{fig:hcst}. A more detailed description of the HCST testbed and the optical components designed and incorporated can be found in Llop-Sayson et al. 2020a \cite{Llop-Sayson_SPIE2020} and 2020b\cite{AVC2020}. The star was simulated using a supercontinuum white-light laser source (NKT Photonics SuperK EXTREME) followed by a tunable single-line filter (NKT Photonics SuperK VARIA) to isolate the specific wavelengths desired for testing. We sampled wavelengths by moving the tunable filter to select narrowbands between 685~nm and 825~nm at increments of 15~nm. At the pupil plane, the beam passes through a filter wheel, either set to a neutral density (ND) filter  (Thorlabs NE20B, OD = 2.0) or an empty slot. Next the adaptive optics system consists of a Boston Micromachines Corporation kilo-DM that controls the wavefront. The deformable mirror has a continuous surface membrane with 34 $\times$ 34 actuators with an inter-actuator separation of 300 $\mu$m. After that, the light is focused and in this setup the focal plane mask used was the scalar vortex mask. Next, a Lyot stop, which is a circular aperture, acts to block $\sim93\%$ of the radius of the 16.4~mm beam. Lastly a field stop is introduced to help block any back reflections before the camera. The remaining light is imaged with a $\sim$ f/50 beam onto the camera (Oxford Instruments Andor Neo 5.5). 

\section{ANALYSIS AND RESULTS}
\label{sec:results}

\subsection{Raw Contrast Results}

\begin{figure} [ht]
\begin{center}
\begin{tabular}{c c} 
\includegraphics[height=5.1cm]{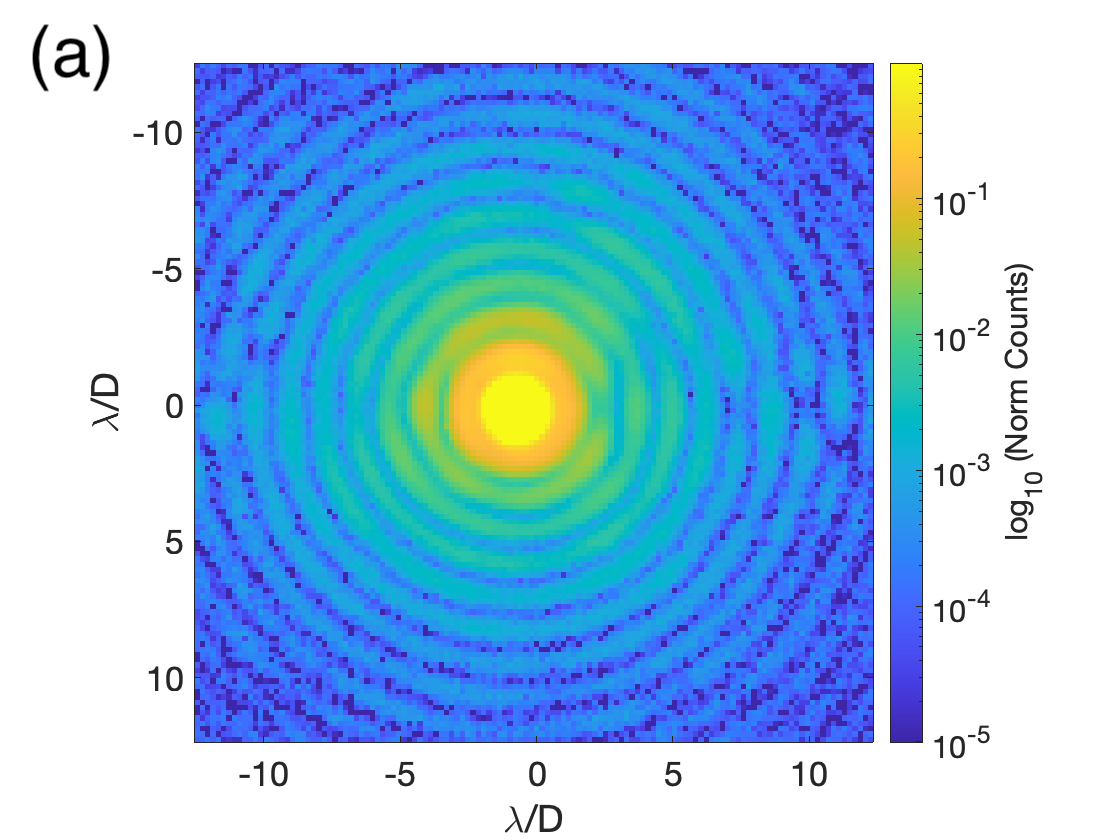}
\includegraphics[height=5.1cm]{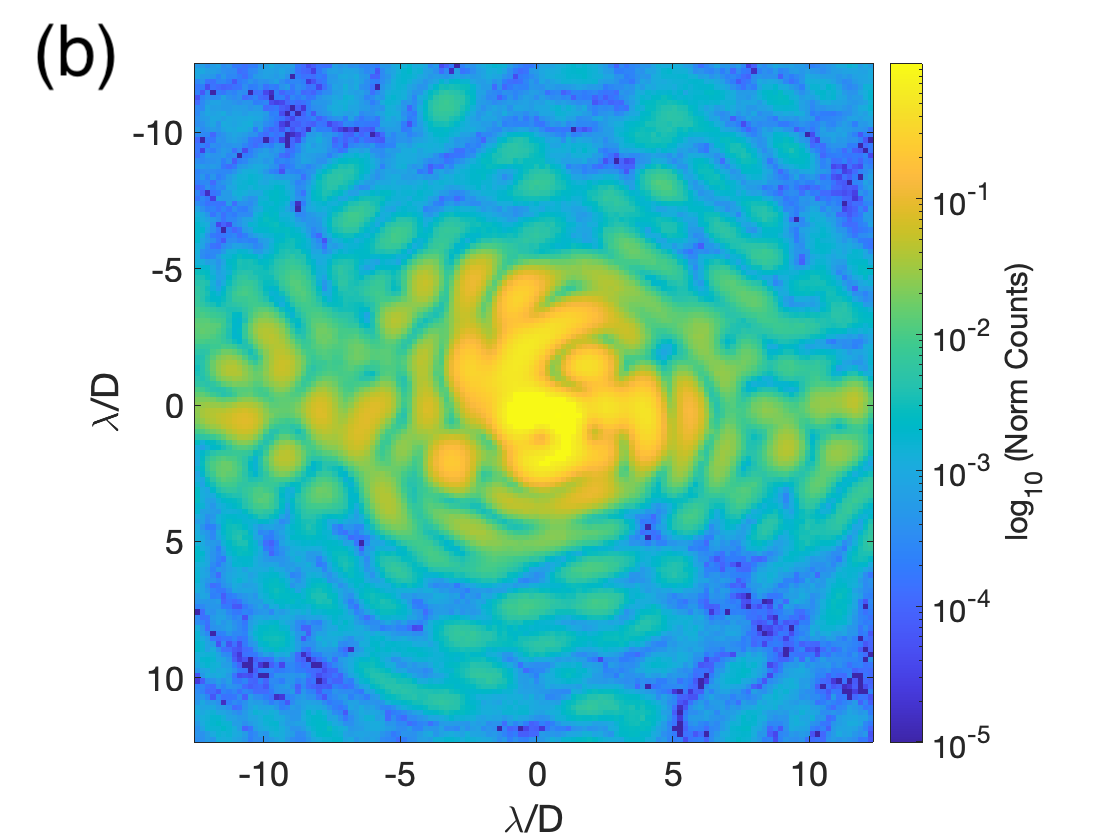}
\end{tabular}
\end{center}
\caption[fig:fpimages] 
{ \label{fig:fpimages} 
(a) Focal plane PSF image of the pseudo-star without SVC mask, taken with a neutral density filter and a 0.001 second integration time. (b) Focal plane image with SVC mask taken with a 0.1 second integration time.}
\end{figure}

\begin{figure} [ht]
\begin{center}
\begin{tabular}{cc} 
\includegraphics[height=4cm]{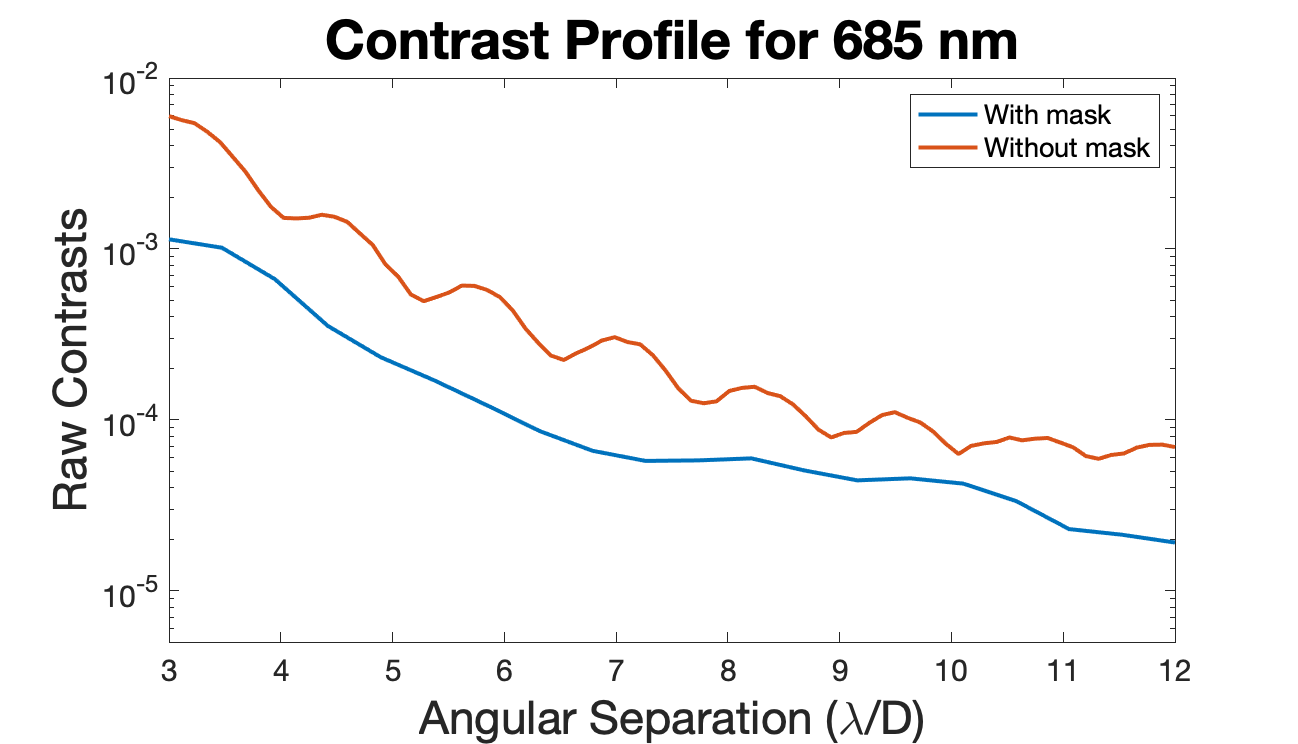} &
\includegraphics[height=4cm]{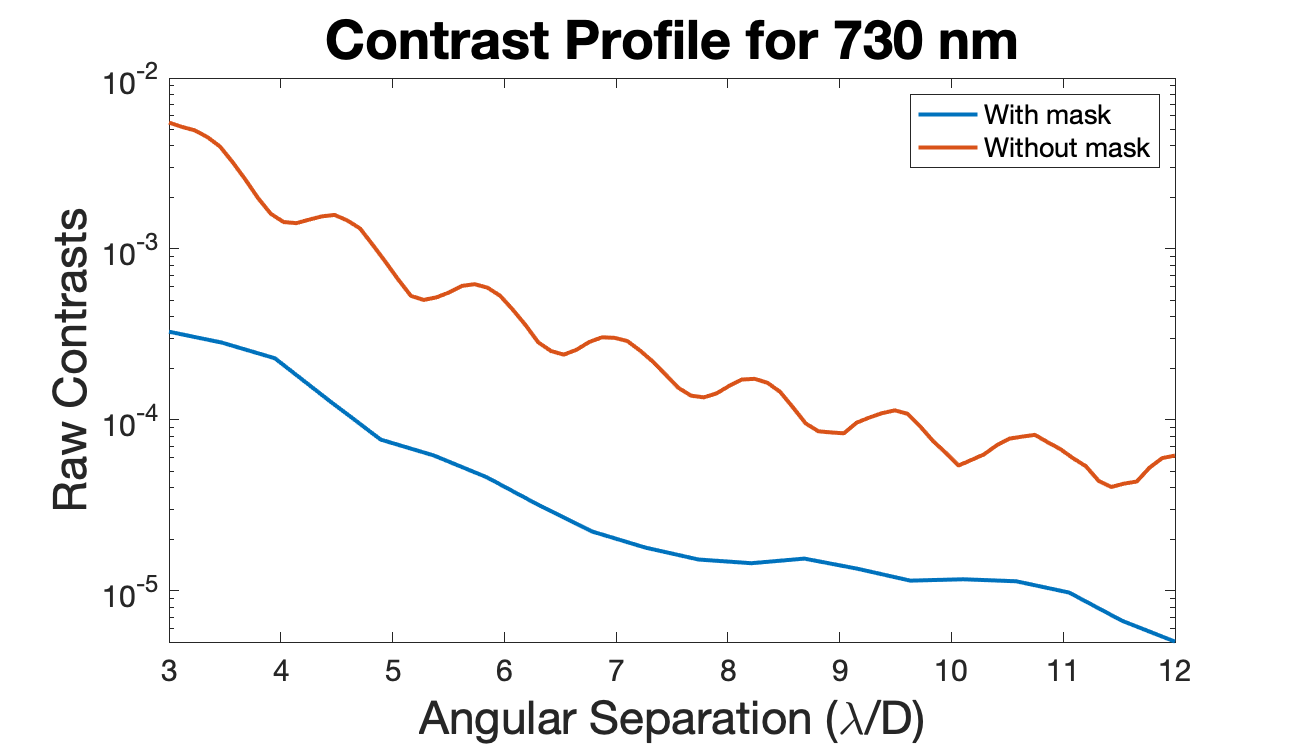} \\
\includegraphics[height=4cm]{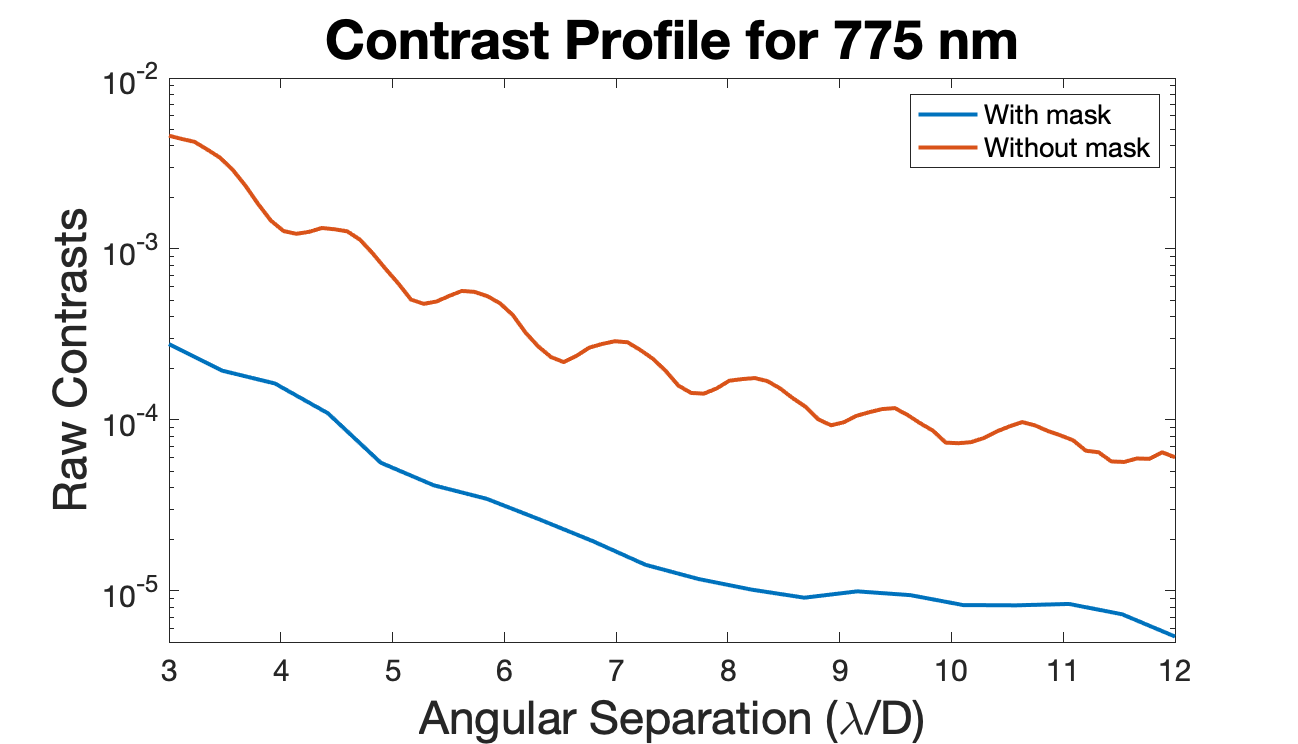} &
\includegraphics[height=4cm]{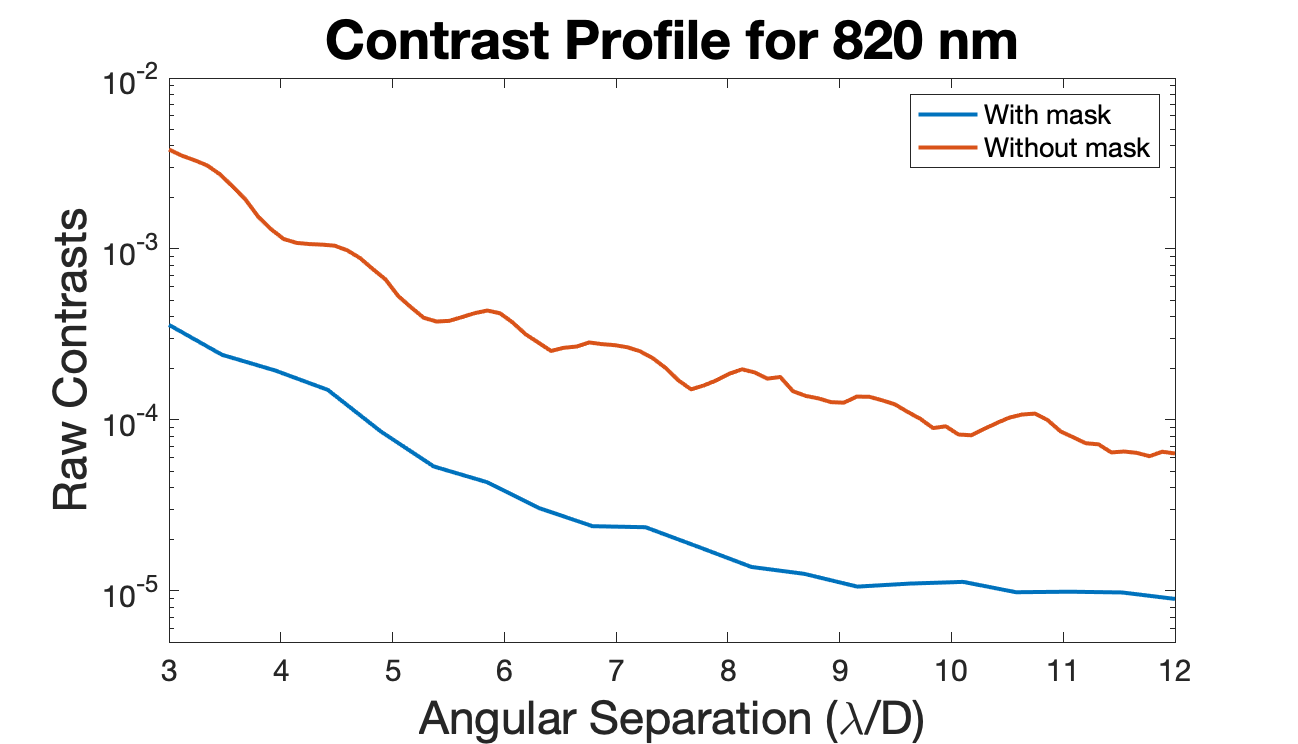}
\end{tabular}
\end{center}
\caption[fig:contrastprofs] 
{ \label{fig:contrastprofs} 
Raw PSF (red) and coronagraphic (blue) contrast profiles from focal plane images for 685 nm, 720 nm, 775 nm, and 820 nm for angular separations from 3 to 12 $\lambda$/D. Overall the scalar vortex coronagraph improved the raw contrasts and suppressed the Airy ring diffraction pattern shown by the bumps in the red PSF curves. }
\end{figure}

\begin{figure} [ht]
\begin{center}
\begin{tabular}{c} 
\includegraphics[height=7.5cm]{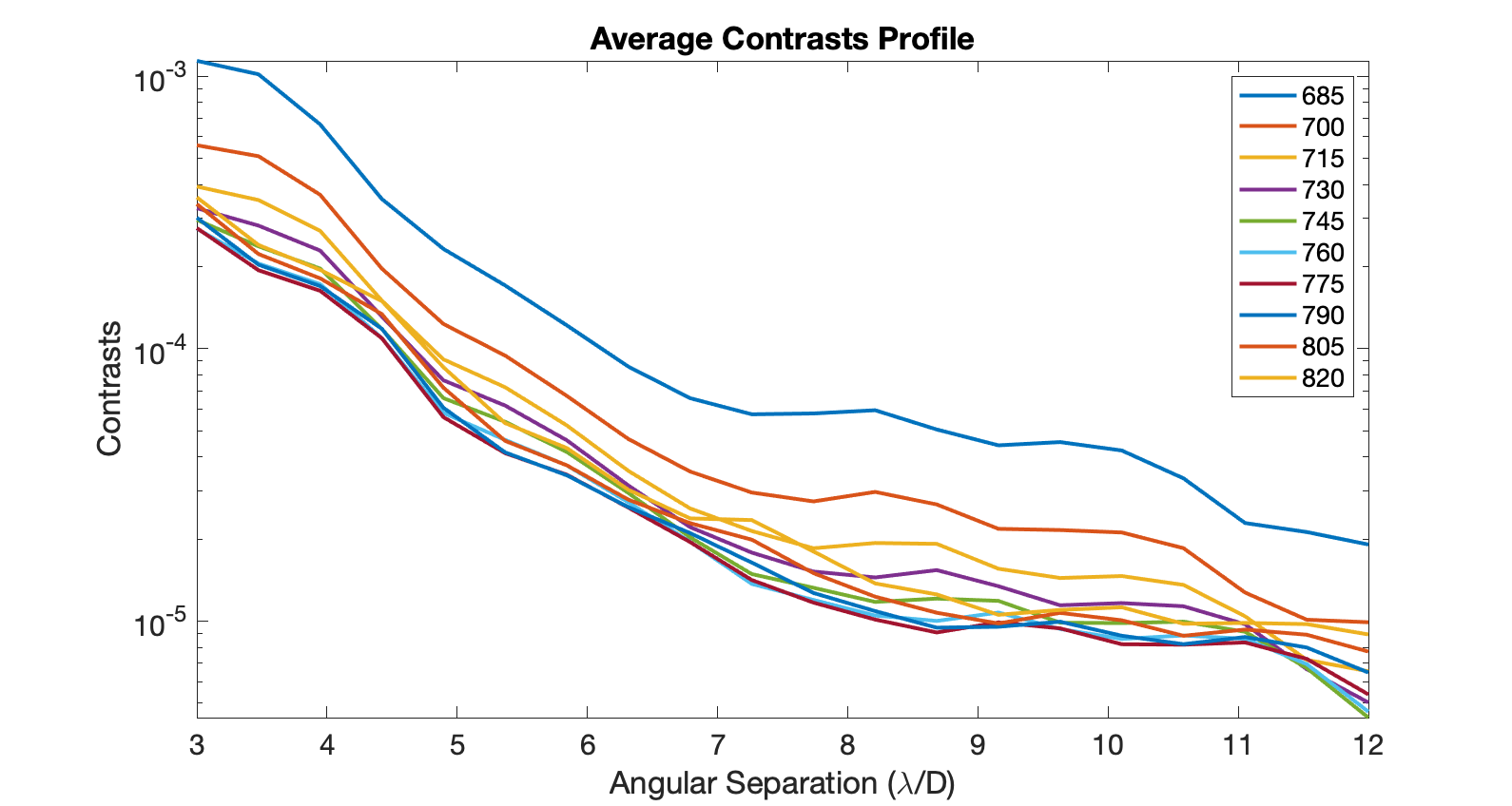}
\end{tabular}
\end{center}
\caption[fig:multicontrasts] 
{ \label{fig:multicontrasts} 
Classical contrast curves plotting raw contrast versus angular separation for wavelengths from 685 nm to 820 nm. These were calculated by taking the ratio of the azimuthal averages at each angular separation to the peak intensity of the unocculted PSF. The best performance is observed at 775~nm (the dark red curve), which was the central designed wavelength for this scalar vortex mask.}
\end{figure}

Focal plane images with and without the scalar vortex mask show qualitatively how well the SVC suppresses starlight. The raw point spread function (PSF) in Fig.~\ref{fig:fpimages} (a) displays a strong Airy ring diffraction pattern all the way out past 12 $\lambda$/D. The coronagraphic image in Fig.~\ref{fig:fpimages} (b) shows that the diffraction pattern is effectively suppressed and a speckle pattern results. Note the raw PSF data was taken with an ND filter in the optical path to avoid saturation of the central pixels. Because of this, the image here was taken with a relatively long integration time of 0.1 seconds. We can compare this to the coronagraphic image to the right which did not require any filter and was only integrated over 0.001 seconds to achieve similar normalized counts as seen in the logarithmic color bars. A further analysis of focal plane images like this in Fig. \ref{fig:contrastprofs} and Fig. \ref{fig:multicontrasts} demonstrate quantitatively the SVC's overall ability to suppress starlight.

\begin{figure} [ht]
\begin{center}
\begin{tabular}{c} 
\includegraphics[height=4.6cm]{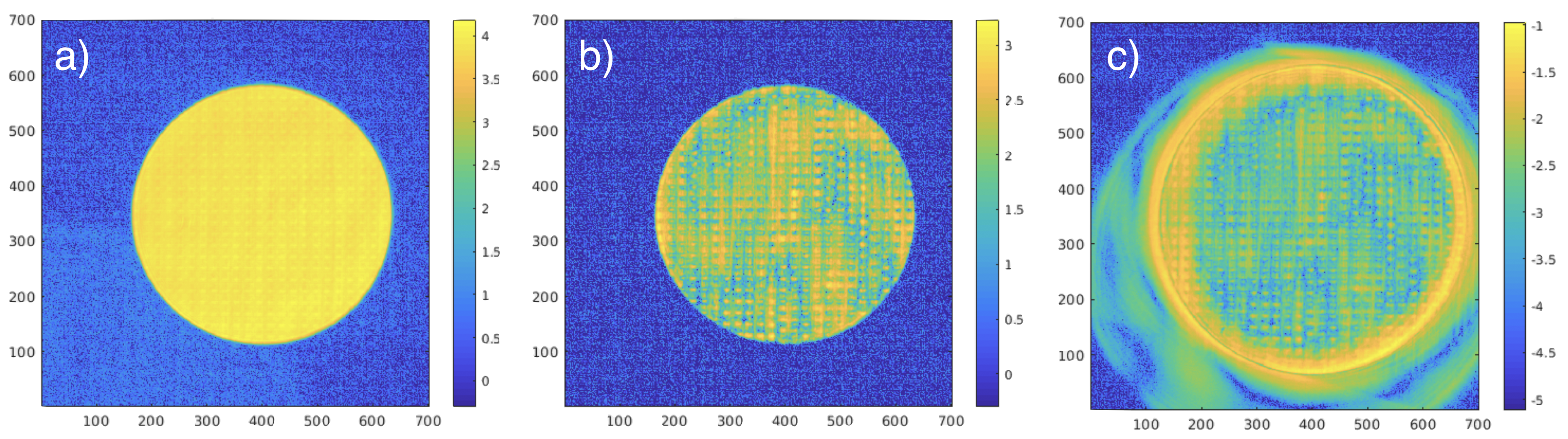}
\end{tabular}
\end{center}
\caption[fig:ppimages] 
{ \label{fig:ppimages} 
(a) Pupil plane image without SVC mask. (b) Pupil plane image with SVC mask. (c) Pupil plane image with SVC mask without Lyot stop. All three pupil images were taken with 0.001 second integration time.}
\end{figure}

In addition to focal plane images, we took pupil plane images to further demonstrate the starlight suppression ability of the SVC. Fig.~\ref{fig:ppimages} shows three pupil plane images at 775nm and all with a 0.001 second integration time: (a) without SVC mask, (b) with SVC mask, and (c) with SVC mask but without the Lyot stop. From this image, we can see the clear “ring of fire” visible along the edges of the pupil plane, which helps to clearly demonstrate how well the SVC is suppressing starlight in the center of the pupil plane. The light at the edge of the pupil is a direct indication that the scalar vortex mask is successfully rejecting light.

To quantitatively characterize the focal plane image starlight suppression, the classical contrast curves in Fig.~\ref{fig:contrastprofs} show the average azimuthal profiles at four wavelengths: 685~nm, 730~nm, 775~nm, and 830~nm with contrast on the y-axis and angular separations stepping from 3 $\lambda$/D out to 12 $\lambda$/D along the x-axis. Here, raw contrast is defined as the averaged intensity of the coronagraphic image divided by the peak intensity of the unocculted pseudo-star image. In each plot the red curve is the contrast profile without the scalar vortex mask and the distinct bumps each correspond to the Airy rings of the diffraction pattern in the PSF. The blue curves are the average contrast profiles with the scalar vortex mask. The overall lower contrast and the lack of bumps in the blue curve shows the SVC’s ability to suppress diffraction. Of these four profiles, it is also clear that the largest contrast improvement is with the 775 nm light, which was the designed central wavelength for these scalar vortex masks.

By sampling over a large range of wavelengths and taking monochromatic measurements at each, we observed the light suppression capability of the SVC was found to degrade as the incoming starlight's wavelength deviated from the central 775 nm, as seen by the dark red curve in Figure~\ref{fig:multicontrasts}.  Furthermore from this graph, it can be seen that observing approximately a 10\% bandwidth (the light blue curve at 760~nm and the dark blue curve at 790~nm) only degrades the contrast by a factor of 2 or 3.

\begin{figure} [ht]
\begin{center}
\begin{tabular}{c c} 
\includegraphics[height=5.4cm]{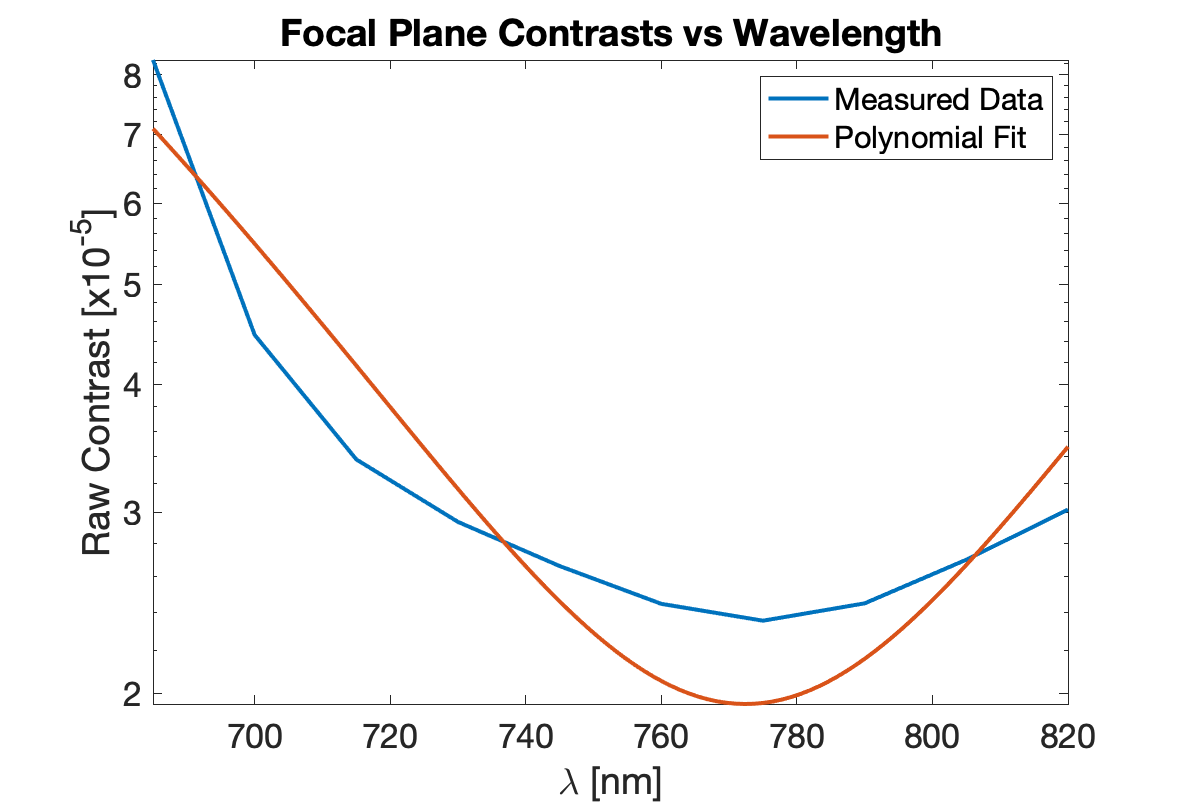} &
\includegraphics[height=5.4cm]{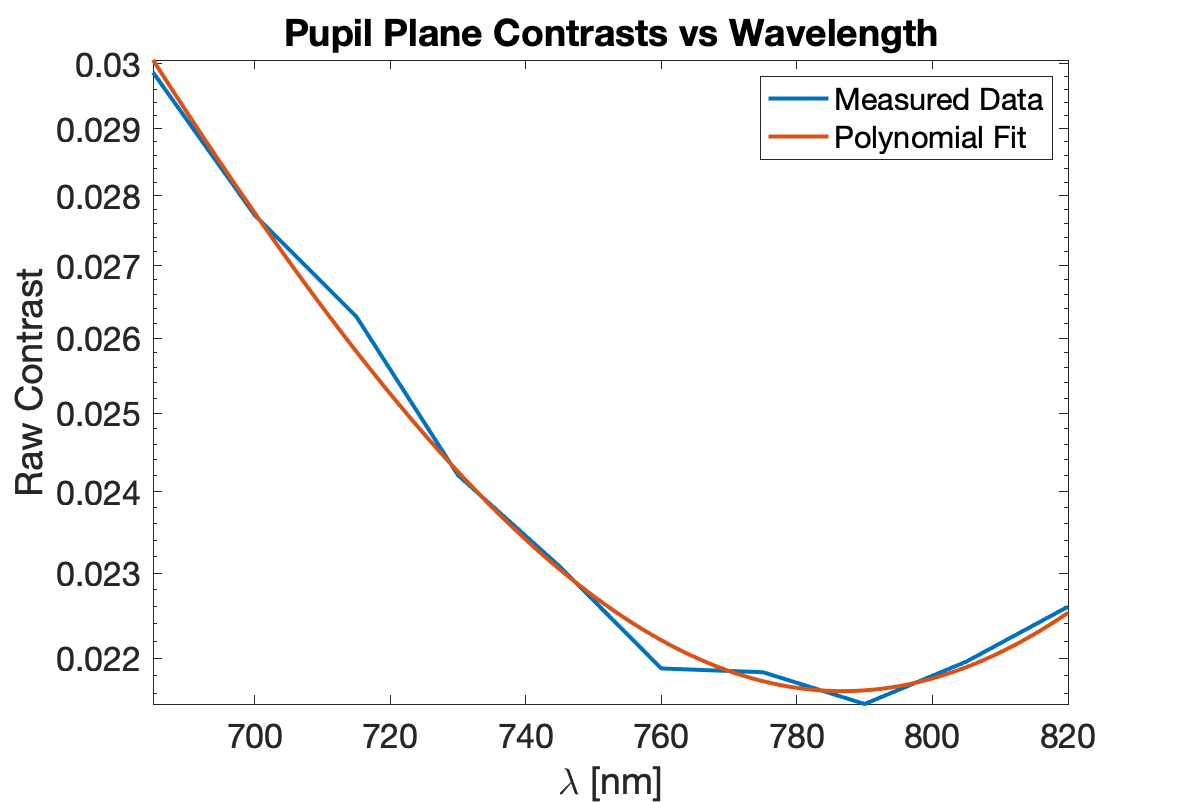}
\end{tabular}
\end{center}
\caption[fig:focpupcontrasts] 
{ \label{fig:focpupcontrasts} 
(a)A focal plane raw contrast vs wavelength curve (blue) averaged across the region from 6 to 8 $\lambda$/D and a quadratic fit (red) to that data. (b) A pupil plane raw contrast versus wavelength curve (blue), which takes the ratio of the flux inside the pupil to the flux outside the pupil, and quadratic fit (red).}
\end{figure}

The focal plane contrast curve in Fig.~\ref{fig:focpupcontrasts} shows the generally monochromatic behavior of the SVC, with peak performance close to 775 nm, the designed central wavelength for this mask. The pupil plane contrast curve in Fig.~\ref{fig:focpupcontrasts} demonstrates in blue the ratio of the total flux inside the pupil vs outside the pupil. In red the parabolic fit of this curve describes a second order dependence on wavelength, as expected from theory\cite{Ruane2019}. The pupil plane data is expected to be more trustworthy because it integrates across all spatial frequencies and is less limited by the speckle patterns in the focal plane. 

\subsection{Preliminary Wavefront Control Results}

After observing some initial success at the raw contrast results, our next step was to attempt improving the contrast further with wavefront control through the deformable mirror. We performed some preliminary wavefront sensing and control (WFSC) with electric field conjugation\cite{Give'On2009} to further suppress residual starlight and to effectively dig a dark hole in some region around the simulated star. For EFC, we used the Fast Linear Least-Squares
Coronagraph Optimization\cite{Riggs2018} (FALCO) software package\footnote{\url{https://github.com/ajeldorado/falco-matlab}}, which is the same the toolbox used to run the HCST.

\begin{figure} [ht]
\begin{center}
\begin{tabular}{c} 
\includegraphics[height=5cm]{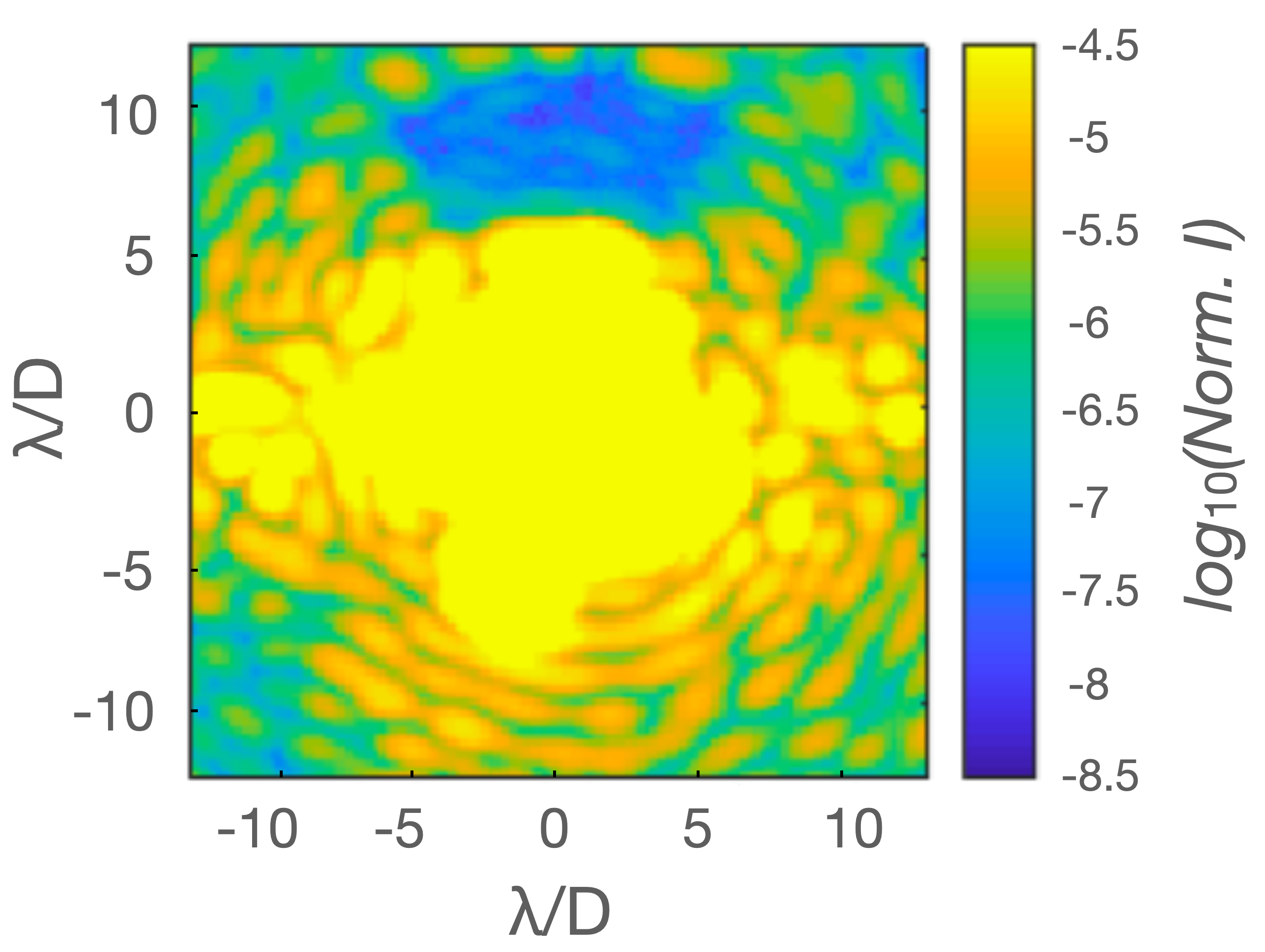}
\end{tabular}
\end{center}
\caption[fig:efc] 
{ \label{fig:efc} 
First attempts at wavefront sensing and control to dig a dark hole resulted in this stellar PSF with a 60 degree dark hole between 6-10 $\lambda$/D, achieving an average dark hole contrast of $10^{-7}$ across the dark hole.}
\end{figure}

Fig.~\ref{fig:efc} shows the stellar PSF with a 60 degree dark hole dug in the range of 6 to 10 $\lambda$/D. It took over 100 iterations to achieve an average dark hole contrast of $10^{-7}$ for a 1\% bandwidth. Although this result indicates some promise for EFC with scalar vortex masks, EFC with the vector vortex mask in the focal plane has not only been able to achieve better contrasts overall, but it's also been able to do it for broadband light. Dark holes dug with VVCs have consistently reached approximately 4e-8 for a 15\% bandwidth for the vector vortex mask in this same HCST laboratory setup in only about 40 iterations.

Examining the progression of EFC on the SVC, we observed the algorithm converges at a floor that is ten times higher than the order of the contrasts achieved by the VVC. Furthermore, we believe that the disproportionately high number of iterations required for the EFC algorithm to converge indicates either the existence of defects in the mask, or a model mismatch between the mask model in FALCO and the physical SVC mask in the bench. We expect that an improved mask model which matches the stepped scalar vortex mask's phase pattern more closely might significantly improve EFC and thereby notably reduce the contrast achievable in the dark hole.

\section{SUMMARY}
\label{sec:conc}

We have demonstrated in this study the successful starlight suppression capabilities of an SVC with a charge 6 stepped scalar vortex focal plane mask. We achieved raw contrasts on the order of $10^{-5}$ from 7 to 9 $\lambda$/D and some preliminary wavefront control with EFC yielded contrasts on the order of $10^{-7}$ from 6 to 10 $\lambda$/D. This study presents experimental evidence that scalar vortex coronagraphs are effective at monochromatic performance and are a worthy avenue of further research for high contrast exoplanet imaging.

\section{FUTURE WORK}
\label{sec:future}
The next phase of research on SVCs is to first improve the mathematical model of the mask used for EFC. For the preliminary wavefront control presented in this paper, the mask model was not updated from the vector vortex mask, and yet the EFC algorithm was still able to converge. With the right mathematical representation of the scalar mask in the model, we expect the EFC algorithm to converge in a dramatically shorter number of iterations and hope it achieves an even better contrast than $10^{-7}$.

Once EFC is improved with a single layer scalar vortex mask, our ultimate goal is to achieve achromatic performance. However an intermediate step toward that goal will be to test other kinds of scalar vortex masks, namely those made by two alternate fabrication methods: ultrafast laser inscription (ULI) and nano-post technologies. We have already secured a few of these other prototype scalar focal plane masks and we are in contact with the Kavli Institute at Caltech for further collaboration on this front.

Efforts toward characterizing the starlight suppression capabilities of each of these types of scalar masks will have the larger objective of figuring out which mask types are viable options for stacking into an achromatic SVC as theorized by early literature \cite{Swartzlander2006}. Studies into designs compatible for layering as well as mechanisms for cementing two masks together without disrupting their critical spiral phase imprinting property will be a major step toward pushing scalar vortex coronagraphs into the achromatic domain.

\acknowledgments 
 This work was supported by the NASA ROSES APRA program, grant NM0018F610. Part of this research was carried out at the Jet Propulsion Laboratory, California Institute of Technology, under a contract with the National Aeronautics and Space Administration.

\bibliography{main} 
\bibliographystyle{spiebib} 

\end{document}